\documentclass[prd,a4paper,nofootinbib,
]{revtex4}

\usepackage{axodraw}
\usepackage{pstricks}
\usepackage{graphicx}

\begin{document}
\title{Embedding $A_4$ into left-right flavor symmetry:\\
Tribimaximal neutrino mixing and fermion hierarchy}

\author{F.~Bazzocchi, S.~Morisi}
\affiliation{
    Instituto de F\'{\i}sica Corpuscular --
  C.S.I.C./Universitat de Val{\`e}ncia \\
  Campus de Paterna, Apt 22085,
  E--46071 Val{\`e}ncia, Spain}
\author{M.~Picariello}
\affiliation{
    Dipartimento di Fisica, Universit\`a del Salento {\em and} Istituto Nazionale di Fisica Nucleare\\
  Via Arnesano, ex collegio Fiorini, I--73100 Lecce, Italy
}
\begin{flushright}
{IFIC/07-62}
\end{flushright}

\begin{abstract}
We address two fundamental aspects of flavor physics:
the mass hierarchy and the large lepton mixing angles.
On one side, left-right flavor symmetry realizes the democratic mass
matrix patterns and explains why one family is much heavier than
the others.
On the other side, discrete flavor symmetry such as $A_4$ leads to the
observed tribimaximal mixing for the leptons.
We show that, by explicitly breaking the left-right flavor symmetry
into the diagonal $A_4$, it is possible to explain both the observed
charged fermion mass hierarchies and quark and lepton mixing angles.
In particular we predict a heavy 3rd family, the tribimaximal mixing
for the leptons, and we suggest a possible origin of the Cabibbo and
other mixing angles for the quarks.
\end{abstract}

\maketitle

\section{introduction}
The recent experimental developments in neutrino physics allowed us to
intensify the studies of the flavor structure of the Standard Model (SM)
and its extensions.
The hardest task was the understanding of the relation between the mass
hierarchies and the large lepton mixing angle between the 2nd and 3rd 
family.
In particular the no-go theorem \cite{Feruglio:2004gu} shows that 
contrary to expectations, a maximal mixing angle $\theta_{23}$
can never arise in the symmetric limit of whatever flavor symmetry
(global or local, continuous or discrete), provided that such a symmetry 
also explains the hierarchy among the fermion masses and is only broken 
by small effects, as we expect for a meaningful symmetry.
A milestone in these studies has been the discovery that mass
hierarchies and mixing angles can be not directly correlated among them
in the flavor symmetry breaking \cite{Picariello:2007yn,Picariello:2006sp}.
In particular, while the mass hierarchies are in general obtained by
using continuous flavor symmetries, such as non-Abelian or $U(1)$ flavor 
symmetry {\'a la} Froggatt-Nielsen, the neutrino experimental data
indicate that the lepton mixing angles may be explained  by discrete
flavor symmetries.
This complementarity between hierarchy and mixing angles allow us to 
escape from the hypothesis of the theorem previously outlined
\cite{Ma:2006wm,Morisi:2007ft}.
The idea is that the flavor symmetry that predicts a large mass for
the 3rd family does not make any prediction on the mixing angles.
However, once the symmetry is broken into a discrete one, then the 
mixing angles are naturally generated.
Another guideline in flavor physics is given by the unification of the
gauge groups. This ingredient forces the field transformations under the
flavor symmetry to be related among them, and strongly reduce the 
degrees of freedom in the model building.

The finite group of even permutations of 4 objects, $A_4$, is the 
smaller non-abelian finite group that contains a triplet irreducible 
representation. It is the first alternating group that is not isomorphic 
to any modulo n group, $Z_n$, or to any direct product of permutation 
groups, $S_n$.
It has been used in the last years 
\cite{Babu:2002dz,Ma:2004yx,Altarelli:2005yp,Feruglio:2007uu,Zee:2005ut,Babu:2005se,Hirsch:2007kh,He:2006dk,Bazzocchi:2007na,Ma:2007ia} to build a 
huge number of models that predict for the lepton sector the 
tribimaximal mixing matrix \cite{Harrison:2002er} with maximal 
atmospheric angle \cite{Grimus:2003wq,Chen:2007zj}, $\theta_{13}=0$
\cite{Duret:2007ux} and $\sin^2 \theta_{sol}=1/3$ 
\cite{Chen:2007af,Chen:2007js,Luhn:2007sy,NimaiSingh:2007zb,Luhn:2007yr}
that agree with neutrino oscillation data.
In \cite{Morisi:2007ft} a non-supersymmetric $SO(10) \times A_4$ grand
unified model, which successfully preserves tri-bimaximal leptonic mixing
and can accommodate all known fermion masses, has been discussed.

In this paper we will show how the embedding of the discrete group $A_4$
into a left-right symmetry allows us to explain the large hierarchy 
between the 3rd and the first two families of quarks and
charged leptons.
At the same time the charged fermion masses of two light families,
that of the neutrinos, and the fermion mixing matrices are related
to the explicitly breaking of the left-right symmetry into the
diagonal $A_4$ and are generated when $A_4$ is spontaneously broken.  
In particular the Cabibbo angle in the quark sector is induced
by higher order operators that explicitly break $SO(3)_L\times SO(3)_R$ 
but preserve the diagonal $A_4$.
Our final aim would be to introduce a gauge unification group
$SO(10)$-like. Since in $SO(10)$ all the Standard Model (SM)
matter fields of one family  belong to the same multiplet, namely
a {\bf 16}-plet, as starting point we will consider a model based
on the discrete flavor symmetry $A_4$ in which left-handed and
right-handed fermions belong to the same representation of $A_4$.

The group $A_4$ has four irreducible representations, three singlets
${\bf 1}$,  ${\bf 1'}$,  ${\bf 1''}$ and a triplet ${\bf 3}$.
Several extensions of the SM are presents in the literature, depending
on the $A_4$ family symmetry realization and the assignments for 
left-handed and right-handed fermion fields.
As we motivated before, we are interested in a realization where both 
left-handed and right-handed fields have the same $A_4$ assignment,
in such a way to be able to perform an embedding into a gauge
grand unified group like $SO(10)$.
Therefore in this paper we will consider a model similar to that
proposed in \cite{Ma:2006wm,Morisi:2007ft} where both left-handed and 
right-handed fields are in the triplet representation of $A_4$.

\section{Mass of the 3rd family from the left-right flavor symmetry} 
\label{sec:top}

The study of models based on the flavor symmetry $U(3)_L\times U(3)_R$
\cite{Kaus:1990ij} or its subgroups both continuos
\cite{Tanimoto:1999pj} or discrete 
\cite{Fritzsch:1989qm,Fritzsch:1994yx,Fritzsch:1998xs} has a long 
history. 
Usually, by imposing a discrete symmetry like $S_{3_L}\times S_{3_R}$,
the charged fermion mass matrix obtained is the so-called democratic 
mass matrix \cite{Harari:1978yi} given by
\begin{equation}
M_{0f}=\frac{m_f}{3}\left(
\begin{array}{ccc}
1&1&1\\
1&1&1\\
1&1&1
\end{array}
\right)\,.
\end{equation}
This matrix has only one eigenvalues different from zero, $m_f$, and can 
be assumed to be the mass  of the 3rd family.
The unitary matrix that diagonalizes the symmetric matrix $M_{0f}$ has 
one angle and the three phases undeterminated. One possible 
parametrization is given by
\begin{equation}
\label{eq:Uparam}
U=
\frac{1}{\sqrt{3}}\,\left(
\begin{array}{ccc}
 \sqrt{2}\cos\theta\,e^{i \alpha}
&\sqrt{2}\sin\theta e^{i (\beta+\gamma)}
&1\\
-e^{i\alpha}(\frac{\cos\theta}{\sqrt{2}}+
             \sqrt{\frac{3}{2}} \sin\theta e^{-i\gamma})
&e^{i \beta}(\sqrt{\frac{3}{2}}\cos\theta
             -\frac{1}{\sqrt{2}}\sin\theta e^{i \gamma})
&1\\
-e^{i \alpha}(\frac{\cos\theta}{\sqrt{2}}
             -\sqrt{\frac{3}{2}}\sin\theta e^{-i\gamma})
&-e^{i \beta}(\sqrt{\frac{3}{2}}\cos\theta
              +\frac{1}{\sqrt{2}}\sin\theta e^{i \gamma})
&1
\end{array}
\right)\,.
\end{equation}
The unknow angle and phases are fixed only after breaking  the 
democratic structure of $M_{0f}$ with a small perturbation $\delta M_f$,
i.e. 
$$ M_f=M_{0f}+\delta M_f\,.$$
For instance, in \cite{Fritzsch:2004xc} $\delta M_f$ is given by
\begin{equation}
\delta M_f \sim
\left(
\begin{array}{ccc}
i \delta &    0      &  0 \\
   0     & -i \delta &  0 \\
   0     &    0      &\epsilon
\end{array}
\right)
\end{equation}
and $M_f$ is diagonalized by
$$
U^f=\left(
\begin{array}{ccc}
 1/\sqrt{2}& 1/\sqrt{6}&1/\sqrt{3}\\
-1/\sqrt{2}& 1/\sqrt{6}&1/\sqrt{3}\\
   0       &-2/\sqrt{6}&1/\sqrt{3}
\end{array}
\right)\,,
$$
obtained by substituting $\alpha=\beta=\gamma=0$ and $\theta=\pi/6$
in eq.~(\ref{eq:Uparam}).

The effect of $\delta M_f$ is to give a small mass to the first and 
second family and to fix the mixing angles.
Another feature of the models based on a symmetry that gives democratic 
charged fermion mass matrices is that up and down quarks are 
diagonalized by almost identical matrices and therefore the CKM can be 
fitted to be close to the identity.
Some attempts of including the neutrinos in this kind of models are quite successful and can fit with good agreement the data \cite{Fritzsch:2004xc}.
Nevertheless, models that have a democratic mass matrix for the charged 
fermions and at the same time predict the tribimaximal mixing matrix for 
the leptons are still missing.

In the following we will build a ``supersymmetry inspired'' model, in 
the sense that the scalars and the SM matter fields we introduce belong 
to supermultiplets and the Lagrangian arises by a superpotential.
In the supersymmetric model proposed in \cite{Altarelli:2005yx} the 
correct alignment of the vevs in the lepton sector, that gives the 
tribimaximal mixing matrix, has been successfully obtained.
However it is difficult to obtain the same result in a context that is 
non supersymmetric.  
By the way, in order to focus on the origin of the mass hierarchies
and mixing angle and to make lighter the reading we will report only the 
Yukawa Lagrangian involving the SM-like fields.

\subsection{Explicitly breaking of $SO(3)_L\times SO(3)_R$ into $A_4$}
  
Let's now extend the flavor symmetry and let's think to $A_4$ as a 
discrete subgroup of the continuous global group $SO(3)_L\times SO(3)_R$.
To implement the idea of explaining both the hierarchy and the mixing 
angles by starting with the same flavor symmetry, we use  two kinds of 
symmetry breaking: the explicit one and the spontaneous one.
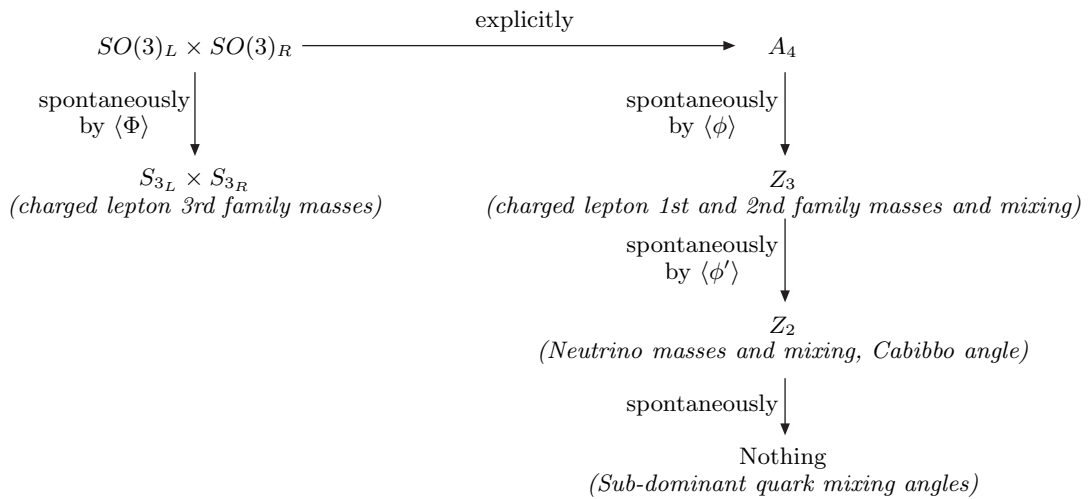
\begin{figure}[bht]
\begin{center}
\begin{picture}(260,200)(0,10)
\Text(0,180)[c]{$SO(3)_L\times SO(3)_R$}
\LongArrow(40,180)(200,180)
\Text(105,190)[l]{explicitly}
\Text(220,180)[c]{$A_4$}
\LongArrow(0,170)(0,140)
\Text(-30,160)[c]{spontaneously}
\Text(-30,150)[c]{by $\langle \Phi \rangle$}
\Text(0,130)[c]{$S_{3_L}\times S_{3_R}$}
\Text(0,120)[c]{{\em (charged lepton 3rd family masses)}}
\LongArrow(220,170)(220,140)
\Text(190,160)[c]{spontaneously}
\Text(190,150)[c]{by $\langle \phi \rangle$}
\Text(220,130)[5]{$Z_3$}
\Text(220,120)[c]{{\em (charged lepton 1st and 2nd family masses and 
mixing)}}
\LongArrow(220,115)(220,85)
\Text(190,105)[c]{spontaneously}
\Text(190,95)[c]{by $\langle \phi' \rangle$}
\Text(220,75)[5]{$Z_2$}
\Text(220,65)[c]{{\em (Neutrino masses and mixing, Cabibbo angle)}}
\LongArrow(220,55)(220,35)
\Text(190,45)[c]{spontaneously}
\Text(220,25)[c]{Nothing}
\Text(220,15)[c]{{\em (Sub-dominant quark mixing angles)}}
\end{picture}
\end{center}
\caption{Diagrammatic representation of the flavor symmetry
structure of the model.
The horizontal arrow indicates the explicit global symmetry breaking
$SO(3)_L\times SO(3)_R\to A_4$ due to the Yukawa terms induced by
a hidden scalar sector. The vertical arrows show the spontaneous
breaking.
The hierarchy among the masses is not directly related to the mixing 
angles.}
\label{fig:symmetries2}
\end{figure}

We impose that the fermion weak doublets $L$, $Q$ transform 
with respect to $SO(3)_L\times SO(3)_R$ as $\sim ({\bf 3},{\bf 1})$ 
while the right-handed fermions $E, U, D$ as
$\sim({\bf 1},{\bf 3})$. 
As explained in \cite{Tanimoto:1999pj}, by assuming  that a discrete 
$S_{3_L}\times S_{3_R}$ is left survived in the spontaneous breaking of 
$SO(3)_L\times SO(3)_R$, the charged fermion mass matrices must have 
the democratic structure.
To write down an invariant term, we introduce a weak scalar singlet 
$\Phi_{ij}\sim({\bf 3},{\bf 3})$  bi-triplet with respect to 
$SO(3)_L\times SO(3)_R$.
The charged fermion masses arise from the following Lagrangian
\begin{equation}\label{eq:L0}
\mathcal{L}_0=
 h^l\,\epsilon_{\alpha\beta}\,H^\alpha_d\,L_i^\beta \,E_j\,
\frac{\Phi_{ij}}{\Lambda}
+h^u\,\epsilon_{\alpha\beta}\,H^\beta_u \,Q_i^\alpha\,U_j\,
\frac{\Phi_{ij}}{\Lambda}
+h^d\,\epsilon_{\alpha\beta}\,H^\alpha_d\,Q_i^\beta \,D_j\,
\frac{\Phi_{ij}}{\Lambda}
\,,
\end{equation}
where $H_u, H_d$ are the scalar components of usual weak doublets
of the MSSM.
The constants $h^{l}$, $h^{u}$, and $h^{d}$ are of order one
while $\Lambda$ is a cut-off.
The $\alpha,\beta$ and $i,j$ are weak and flavor indeces 
respectively, and $\epsilon_{\alpha\beta}$ is the antisymmetric
tensor.
In ref.\cite{Tanimoto:1999pj} has been shown that the minimization of 
a potential, invariant with respect to $SO(3)_L\times SO(3)_R$,
leads $S_{3_L}\times S_{3_R}$ invariant vev. 
The scalar field $\Phi_{ij}$ will take its vev in the direction
\begin{equation}
\label{eq:massdem}
\langle \Phi \rangle \propto \left(
\begin{array}{ccc}
1&1&1\\
1&1&1\\
1&1&1
\end{array}
\right)\,,
\end{equation}
and the resulting charged fermion masses are the democratic mass matrix.

The masses of the first and second families and that of the
neutrinos arise once we include, in the Yukawa Lagrangian of 
eq.~(\ref{eq:L0}), terms that explicitly break the continuous $SO(3)_L\times SO(3)_R$ but that preserve the discrete diagonal
subgroup $A_4$. 
For example, we can assume the presence of an hidden scalar sector that 
breaks spontaneously  the continuous $ SO(3)_L\times SO(3)_R$ into the 
diagonal $A_4$.
The explicit breaking terms will be of the form 
\begin{eqnarray}\label{eq:L1}
\mathcal{L}_1&=& 
\frac{\delta^l}{\Lambda}
\,\epsilon_{\alpha \beta}\,H_d^\alpha\,(L^\beta \,E\,\phi)+
\frac{\delta^u}{\Lambda}
\,\epsilon_{\alpha \beta}\,H_u^\beta \,(Q^\alpha\,U\,\phi)+ 
\frac{\delta^d}{\Lambda}
\,\epsilon_{\alpha \beta}\,H_d^\alpha\,(Q^\beta \,D\,\phi)\nonumber\\
&+&
\frac{x}{\Lambda}
\,\xi'\,(L^\alpha\,\sigma^n_{\alpha\beta}\,L^\beta)''\,\tilde{\Delta}_n
+ \frac{y}{\Lambda}
\,\left(\phi'\,L^\alpha\,\sigma^n_{\alpha\beta}\,L^\beta\right)
  \Delta_n
\end{eqnarray}
where the fields in eq.~(\ref{eq:L1}) transform according to table 
\ref{tab:1} and $\Lambda$ is the cut-off scale of the model.
The labels $\alpha,\beta$ are again weak indeces, $\sigma^n$ are the Pauli matrices and $\Delta_n, \tilde{\Delta}_n$
are weak triplets, as reported in table \ref{tab:1}.
We have introduced an additional $Z_5$ symmetry that affects only
the scalar sector and avoids the presence of unwanted Yukawa
couplings as done for instance in \cite{Altarelli:2005yp,Bazzocchi:2007na}.
We remember that, if $a=(a_1,a_2,a_3)$ and $b=(b_1,b_2,b_3)$ are two 
$A_4$ triplets, then ${\bf 1}\sim (ab)=(a_1b_1+a_2b_2+a_3b_3)$,
${\bf 1'}\sim (ab)'=(a_1b_1+\omega^2 a_2b_2+\omega a_3b_3)$ and
${\bf 1''}\sim (ab)''=(a_1b_1+\omega a_2b_2+\omega^2 a_3b_3)$ \cite{Morisi:2007ft}. 
\begin{table}[tb]
\begin{center}
\begin{tabular}{|c||c|c|c|c|c|c|c| |c| |c|c|c|c|c|}
\hline
\multicolumn{1}{|c}{}
&\multicolumn{7}{c|}{MSSM fields}
&\multicolumn{1}{c}{}
&\multicolumn{5}{|c|}{Fields of the explicit breaking into $A_4$ }\\
\hline
\multicolumn{1}{|c}{}
&$\quad\hat{L}\quad$  &$\quad\hat{Q}\quad$&$\quad\hat{E}  \quad$
&$\quad\hat{U}\quad$  &$\quad\hat{D}\quad$&$\quad\hat{H}_u\quad$
&$\quad\hat{H}_d\quad$&$\quad\Phi\quad$   &$\quad\hat{\Delta}\quad$&
$\quad\hat{\tilde{\Delta}}\quad$&$\quad\hat{\phi}\quad$&
$\quad\hat{\phi'}\quad$&$\quad\hat{\xi}'\quad$\\
\hline
$\mbox{Weak } SU(2)$& {\bf 2}&{\bf 2}&{\bf 1}&{\bf 1}&{\bf 1}&
{\bf 2}&{\bf 2}&{\bf 1}
&{\bf 3}&{\bf 3}&{\bf 1}&{\bf 1}&{\bf 1}\\
$SO(3)\times SO(3)$&({\bf 3},{\bf 1})&({\bf 3},{\bf 1})
&({\bf 1},{\bf 3}) &({\bf 1},{\bf 3})
&({\bf 1},{\bf 3}) &({\bf 1},{\bf 1})&({\bf 1},{\bf 1})&
({\bf 3},{\bf 3})&
&&&&\\
$A_4$ &{\bf 3}& {\bf 3}&{\bf 3} &{\bf 3} &{\bf 3}&{\bf 1}&{\bf 1}&
&{\bf 1}&{\bf 1}&{\bf 3}&{\bf 3}&${\bf 1'}$\\
$Z_5$&1&1&1&1&1&1&1&1&$\omega_5^2$&$\omega_5^3$&1&$\omega_5^3$&$\omega_5^2$\\
\hline
\end{tabular}
\caption{The supermultiplet content of the model. We have denoted 
$\omega_5$ the discrete charge that satisfy $\omega_5^5=1$.}
\label{tab:1}
\end{center}
\end{table}

Under the hypothesis that the breaking of $SO(3)_L\times SO(3)_R$ into 
$A_4$ happens in a hidden scalar sector and then it is transmitted
to the fermions through the integration of the heavy fields, it is quite 
natural to assume that the explicit breaking terms in eq.~(\ref{eq:L1}), 
to be added to the Lagrangian of eq.~(\ref{eq:L0}), are small. 
To get more familiar with the embedding of $A_4$ into
$SO(3)_L\times SO(3)_R$, we report the decomposition of some 
representations of $SO(3)_L\times SO(3)_R$ into the representations of  
$A_4$ in table \ref{tab:2}.
\begin{table}[b]
\begin{center}
\begin{tabular}{|c|c|c|}
\hline
$\quad SO(3)_L\times SO(3)_R\quad $ &$\quad\quad SO(3)\quad\quad$& $ \quad\quad\quad A_4\quad\quad\quad$\\
\hline
({\bf 3},{\bf 1}) & {\bf 3}&{\bf 3}\\
({\bf 1},{\bf 3}) & {\bf 3}&{\bf 3}\\
({\bf 3},{\bf 3}) & {\bf 1}+{\bf 3}+{\bf 5}&
 {\bf 1}+{\bf 3}+${\bf 1'}$+${\bf 1''}$+${\bf 3'}$\\
\hline
\end{tabular}
\caption{ Decomposition of some representations  of   $SO(3)_L\times SO(3)_R$ into the representations of  $ A_4$.
To clarify the decompositions, we also report the representations under $SO(3)$ diagonal.}
\label{tab:2}
\end{center}
\end{table}
The correspondences for the fundamental representations are obvious.
We can spend few words on the bi-fundamental.
The $({\bf 3},{\bf 3})$ representation of $SO(3)_L\times SO(3)_R$ gives
the irreducible representations ${\bf 1}+{\bf 3}+{\bf 5}$ when
the group is broken to the diagonal $SO(3)$ that in turn give
${\bf 1}+{\bf 3}+{\bf 1'}+{\bf 1''}+{\bf 3'}$ when $SO(3)$
is broken into $A_4$ as explained in \cite{Hagedorn:2006ug}. 

When  $\phi$ takes vev as 
$\langle \phi \rangle=v_\phi\,(1,1,1)$ we have for the charged
leptons
\begin{equation}\label{eq:epsi}
\frac{\delta^l}{\Lambda}\,\epsilon_{\alpha\beta}\,H_d^\alpha 
\,\left(L^\beta\,E\,\phi\right)
=\epsilon_{\alpha\beta}\,H_d^\alpha
\left[\gamma^l_1 \,(L_2^\beta E_3+L_3^\beta E_1+L_1^\beta E_2)+ 
\gamma^l_2 \,(L_3^\beta E_2+L_1^\beta E_3+L_2^\beta E_1)\right]\,,
\end{equation}
with $\gamma^l_i=\delta^l_i v_\phi/\Lambda$ and
the two $\delta_i$ arise by the two different contractions of $A_4$.  
Similar expressions are obtained for the quarks.
The effect of the explicit breaking terms in the mass matrices 
is translated in a perturbation of the democratic mass matrix
of eq.~(\ref{eq:massdem}), that is 
\begin{equation}\label{break1}
 M^f  =\frac{m_3^f}{3}\left(
\begin{array}{ccc}
1&1&1\\
1&1&1\\
1&1&1
\end{array}
\right)
\to
\,\frac{m^f_3}{3}\left(
\begin{array}{ccc}
1&1+\gamma^f_1&1+\gamma^f_2\\
1+\gamma^f_2&1&1+\gamma^f_1\\
1+\gamma^f_1&1+\gamma^f_2&1
\end{array}
\right)\equiv
v_f\, \left(
\begin{array}{ccc}
h_0^f&h_1^f&h_2^f\\
h_2^f&h_0^f&h_1^f\\
h_1^f&h_2^f&h_0^f
\end{array}
\right)\,,
\end{equation}
with the obvious correspondences $m_3^f/3= v_f h_0^f=v_f v_\Phi/\Lambda$,
$v_f h_{1,2}^f=(1+\gamma^f_{1,2})\cdot m_3^f/3$ and $v_f=v_{u,d}$.
The mass matrix of eq.~(\ref{break1}) is diagonalized by
\begin{equation}\label{eq:U}
\tilde{U}_\omega=
\frac{1}{\sqrt{3}}\left(
\begin{array}{ccc}
\omega&\omega^2&1\\
\omega^2&\omega&1\\
1&1&1
\end{array}
\right)\,,
\end{equation}
corresponding to the $U$ of eq.~(\ref{eq:Uparam}) with $\theta=\pi/4$, $\alpha=2 \pi/3$, $\beta=5 \pi/6$ and $\gamma=\pi/2$.
$M^f$ of eq.~(\ref{break1}) gives an heavy 3rd family
mass $m^f_3$ and small 1st and 2nd family masses satisfying
\begin{equation}\label{13}
\frac{m^f_1}{m^f_3}=\frac{\omega \gamma^f_1+\omega^2 \gamma^f_2}{3},\qquad\frac{m^f_2}{m^f_3}=\frac{\omega^2 \gamma^f_1+\omega \gamma^f_2}{3}\,.
\end{equation}

\subsection{Neutrino sector}
\label{sec:newneut}
The Yukawa interactions for the neutrinos are the following
\begin{equation}\label{14}
\mathcal{L}_\nu=
  \frac{x}{\Lambda}\,\xi'\,
(L^\alpha \sigma^n_{\alpha\beta} L^\beta)'' \tilde{\Delta}_n
+ \frac{y}{\Lambda}\,
\left(\phi' L^\alpha \sigma^n_{\alpha\beta} L^\beta\right) \Delta^n
\,,
\end{equation}
where the scalars $\xi'$ and $\phi'$ are  singlets  of the weak
$SU(2)_L$ and transform with respect to $A_4$ as 
${\bf 1'}$ and ${\bf 3}$  respectively.
The scalars  $\Delta$  and $\tilde{\Delta}$  are singlets of $A_4$
and triplets of the weak $SU(2)_L$. 
When the triplet field $\phi'$ takes vev in the $A_4$ direction
$\langle \phi \rangle\sim(0,0,1)$ - notice that this alignment is 
different from the one used in many models as for example in
\cite{Morisi:2007ft,Altarelli:2005yp} -,
the resulting neutrino mass matrix is given by
\begin{equation}\label{15}
M_\nu=\left(
\begin{array}{ccc}
a & b& 0\\
b &\omega\, a & 0\\
0 &0 &\omega^2\, a 
\end{array}
\right)=
\tilde{V}_\nu
\left(
\begin{array}{ccc}
\omega^2\, a +b& 0& 0\\
0 &\omega^2\, a & 0\\
0 &0 &-\omega^2\,a+ b 
\end{array}
\right)\tilde{V}_\nu^T,\qquad
\tilde{V}_\nu=\left(
\begin{array}{ccc}
\frac{\omega}{\sqrt{2}} &0 & \,-i\frac{\omega^2}{\sqrt{2}} \\
\frac{\omega^2}{\sqrt{2}} &0 & i\,\frac{\omega}{\sqrt{2}}\\
0&1&0
\end{array}
\right).
\end{equation}
The charged leptons are diagonalized by $L\to \tilde{U}_\omega\,L$, so we obtain a tribimaximal mixing for the lepton sector, that is 
\begin{equation}\label{16}
V_{PMNS}=\tilde{U}^\dagger\cdot \tilde{V}_\nu=
\left(
\begin{array}{ccc}
\frac{2}{\sqrt{6}} &\frac{1}{\sqrt{3}} &0 \\
-\frac{1}{\sqrt{6}} &\frac{1}{\sqrt{3}} &-\frac{1}{\sqrt{2}} \\
-\frac{1}{\sqrt{6}} &\frac{1}{\sqrt{3}} & \frac{1}{\sqrt{2}}
\end{array}
\right)
\end{equation}
and the
the neutrino masses result to have the same expressions of \cite{Altarelli:2005yp}.

\subsection{The origin of the Cabibbo angle}
In eq.~(\ref{eq:L1}) we have reported the leading  $A_4$ invariant terms that arise after the explicitly breaking of $SO(3)_L\times SO(3)_R$.
We include now the higher order operators suppressed by powers of the cut-off scale  $\Lambda$. The first terms at order $1/\Lambda^2$
that change the structure of the  charged fermion mass matrices above are
\begin{eqnarray}\label{eq:L3}
\mathcal{L}_3& =& 
 g^l \,\epsilon_{\alpha \beta}\,H_d^\alpha\,
\left(L^\beta  E\,\frac{\phi' \xi' }{\Lambda^2}\right)
+g^u \,\epsilon_{\alpha \beta}\,H_u^\beta \,
\left(Q_\alpha U\,\frac{\phi' \xi' }{\Lambda^2}\right)
+g^d \,\epsilon_{\alpha \beta}\,H_d^\alpha\,
\left(Q_\beta  D\,\frac{\phi' \xi' }{\Lambda^2}\right)\,.
\end{eqnarray}
With the inclusion of these contributions the charged fermion mass 
matrices of eq.~(\ref{break1}) become
$$
M^f=v_f\,\left(
\begin{array}{ccc}
h_0^f&h_1^f&h_2^f\\
h_2^f&h_0^f&h_1^f\\
h_1^f&h_2^f&h_0^f
\end{array}
\right)
\to
M^f_{eff}=v_f\,\left(
\begin{array}{ccc}
h_0^f&h_1^f+3\rho^f_1 &h_2^f\\
h_2^f+3\rho^f_2&h_0^f&h_1^f\\
h_1^f&h_2^f&h_0^f
\end{array}
\right)
$$
where $v_f=v_{u,d}$,  $3 \rho^f_i\sim g^f_i  v_{\phi'} v_\xi/\Lambda^2$.
The $g^f_i$, $i=1,2$, arise from the possible different contractions of
${\bf 3}$-plet of $A_4$ to give a singlet ${\bf 1''}$ and the factor
3 is introduced to simplify the subsequent formulas.
In the basis rotated by $\tilde{U}_\omega$ of eq. (\ref{eq:U}),
namely $\tilde{M}^{f}\equiv \tilde{U}_\omega^\dagger\,M^f_{eff}\, \tilde{U}_\omega$, the charged fermion  mass matrices are now approximatively given by
\begin{eqnarray}
 \tilde{M}^{f}&\approx& 
{\tilde{v}_f \,\left(
\begin{array}{ccc}
r^f_1+ \epsilon^f_1\omega +\epsilon^f_2 \omega^2 & \epsilon^f_1+\epsilon^f_2 & \epsilon^f_1 \omega^2 +\epsilon^f_2 \omega \\
 \epsilon^f_1+\epsilon^f_2&r^f_2+\epsilon^f_1\omega^2 +\epsilon^f_2 \omega& \epsilon^f_1+  \omega^2 \epsilon^f_2 \\
\epsilon^f_1 \omega^2 +\epsilon^f_2 \omega &\epsilon^f_1+  \omega^2 \epsilon^f_2 &r^f_3+ \epsilon^f_1+\epsilon^f_2 
\end{array}
\right)}
\end{eqnarray}
where $r^f_i=m^f_i/\tilde{v}_f$, $\tilde{v}_f=v_f v_\Phi/\Lambda$ and 
$\epsilon^f_i=\rho^f_i/\tilde{v}_f$. Let's assume that the
$\epsilon^f_i$ are arbitrary parameters of  $O(\lambda^5)$ , where 
$\lambda$ is the Cabibbo angle. The crucial point is that this 
assumption has the consequences that the higher order operators
give negligible effects in the down and charged lepton sectors, since 
for the down and charged leptons we have $(r_1^{d,l},r_2^{d,l},r_3^{d,l})\sim(\lambda^4,\lambda^2,1)$ and 
$\tilde{M}^{d,l}$ may be considered diagonals. On the contrary for the 
up quarks we have that 
$(r_1^{u},r_2^{u},r_3^{u})\sim(\lambda^7,\lambda^4,1)$ 
and therefore the off-diagonal entries (1,2) and (2,1) cannot
be neglected: the matrix $\tilde{M}^{u}$ is diagonalized by a
rotation in the 12 plane with $\sin\theta_{12}\approx \lambda$.
This rotation produces the Cabibbo angle in the CKM.
In fact while $M^d$ is still diagonalized by $U_\omega$, 
we have that $M^u$ is diagonalized by
$V_L^{u \dag}\, U_\omega^\dagger M^u U_\omega \,V^u_R$ where $V_{LR}^u$ 
are unitary matrix, rotations in the 12 plane, and therefore the CKM 
mixing matrix is  given by
$$
V_{CKM}=V_L^{u^\dagger}\, U_\omega^\dagger\,U_\omega\equiv V_L^{u^\dagger}\,.
$$
The charm and top quark masses are almost unaffected by the corrections and still are given by $\tilde{v}_u r_2^u$ and $\tilde{v}_u r_3^u$ respectively. The up quark mass is obtained by   tuning the $\epsilon^u_i$ and is given more or less by
$$m_u \approx \tilde{v}_u\, (\epsilon^u_1\omega +\epsilon^u_2 \omega^2 ).$$

In  \cite{He:2006dk}  the full CKM was obtained  by breaking the  $Z_2$ 
symmetry that survives when a triplet of $A_4$ takes vev in the 
direction $(1,0,0)$.  In our model  we suggest that the origin of the 
Cabibbo angle is instead in the  $A_4 $ invariant subleading corrections 
to the Yukawa interactions. The breaking of the  residual $Z_2$ symmetry 
allows instead to generate the complete CKM.
The main difference between our model and some previous models, where
the subleading corrections in the charged lepton matrix are too small
to generate a Cabibbo angle in order to keep the lepton mixing angles
inside the bounds given by the experimental data, is
related to the different assignment and the $U(1)$ flavor symmetry one
introduces in order to explain the mass hierarchies.
For example, in \cite{Altarelli:2005yx} the left-handed fields belonged 
to a triplet of $A_4$, while at the right-handed fields was
given the assignment ${\bf 1},{\bf 1''},{\bf 1'}$ and they
have $U(1)$ charges $(2q,q,0)$ where $q$ is a real number. 

\section{Grand unified group $SO(10)\times SU(3)$}
As already explained in the introduction our final aim would be the
construction of a grand unified $SO(10)$-like model.
Let us  assume the group  $A_4$ as flavor
symmetry and the ``constrain'' of assigning right and left-handed
fermion fields to the same representations.
Since $A_4$ has four irreducible representations,
three singlets ${\bf 1}$, ${\bf 1'}$ and ${\bf 1''}$,
and a triplet ${\bf 3}$,
clearly we have just few possibilities.
For example if we assign the three {\bf 16}-plets to
${\bf 1}$, ${\bf 1'}$ and ${\bf 1''}$ we obtain a mass
matrix for the charged fermions of the form
\begin{equation}\label{1}
M_f=\left(
\begin{array}{ccc}
\alpha&0&0\\
0&0&\beta\\
0&\beta&0
\end{array}
\right)\,,
\end{equation}
where $\alpha$ and $\beta$ are arbitrary parameters, that gives for 
instance the wrong prediction $m_c=m_t$. The situation is better only
when the three {\bf 16}-plets transform as a triplet of $A_4$. Indeed, 
it has been showed in \cite{Morisi:2007ft} that the assignment of both 
left-handed and right-handed SM fields to triplets of $A_4$, that is 
therefore compatible with  $SO(10)$,  can be lead to  the 
charged fermion  textures proposed by E.Ma \cite{Ma:2006wm} and  given 
by 
\begin{equation}\label{2}
M_f=\left(
\begin{array}{ccc}
h^f_0 & h^f_1& h^f_2\\
h^f_2 & h^f_0 & h^f_1 \\
h^f_1 & h^f_2&h^f_0
\end{array}
\right)\,,
\end{equation}
with $h^f_0,\,h^f_1$ and $h^f_2$  distinct parameters. In 
\cite{Morisi:2007ft}, in order to obtain a mass matrix of the form of 
$M_f$ in eq.~(\ref{2}) without spoiling the predictions of the neutrino 
sector,   higher order operators were introduced containing
simultaneously the $SO(10)$ representations ${\bf45}_{T3R}$ and 
${\bf45}_{Y}$ that took vevs in the isospin and hypercharge
directions respectively.
A renormalizable $SO(10)\times A_4$ model has been  recently studied in \cite{Grimus:2007tm} where however the $A_4$ flavor symmetry does 
not enforce a tribimaximal mixing in the lepton sector.

The group $SO(3)_L\times SO(3)_R$ is not compatible with $SO(10)$ since the  {\bf 16}-plet contains both left-handed and right-handed fields that
belong to different representations of  $SO(3)_L\times SO(3)_R$. 
We have therefore to search for a  continuous  group larger than $SO(3)_L\times SO(3)_R$, with rank bigger
than $2+2=4$, and that has a triplet as fundamental representation. 
The group $SU(3)$ seems us a good candidate. 
The scalar field $\Phi_{ij}\sim ({\bf 3},{\bf 3})$ of the model we have just considered will correspond to the ${\bf \bar{6}}$ representation of $SU(3)$ whose  vev is compatible with the democratic mass matrix.

Without entering into the details of the realization of an $SO(10)\times 
SU(3)$ model 
\cite{Antusch:2007re,Koide:2007sr,King:2003rf,King:2001uz,Chkareuli:2001dq}
that we leave for a future work,  we want to suggest how
its realization could be achieved using non renormalizable operators.
We can think that such operators  arise by integrating out
some heavy extra fermions that are coupled to the matter fields at the 
renormalizable level, for instance see 
\cite{Malinsky:2007qy,McKeen:2007ry,Berezhiani:1996bv,Barr:2007ma}.
The effective $SO(10)$ invariant Lagrangian could be
$$
\mathcal{L}=\mathcal{L}_{SU(3)}+\delta \mathcal{L}_{A_4}\,,
$$
where $\mathcal{L}_{SU(3)}$ is $SO(10)\times SU(3)$ invariant and 
$\delta \mathcal{L}_{A_4}$ is the explicit breaking term of the $SU(3)$ 
symmetry that, at this level, leaves $SO(10)$ unbroken.
In particular the $SU(3)$ invariant term is
$$
\mathcal{L}_{SU(3)}=\lambda\, {\bf 16}\,{\bf 16}\,{\bf 10}_D\,{\bf 45}_{T_{3R}}\,{\bf 45}_Y\,,
$$
where ${\bf 10}_D$ transforms as ({\bf 10},$\overline{\bf 6}$) with respect to $SO(10)\times SU(3)$. The scalar fields
${\bf 45}_{T_{3R}}$ and ${\bf 45}_Y$ are singlets of $SU(3)$ and their vevs are proportional  to the right-handed
isospin and to the hypercharge respectively. 
Thanks to the ${\bf 45}_{T_{3R}}$ and ${\bf 45}_Y$ scalar fields,
the above operator does not give any contribution to the neutrino 
sector, while all the charged fermion mass matrices are of democratic 
form if ${\bf10}_D$ takes vev along the direction that preserves a
$S_3\times S_3$ subgroup of $SU(3)$.
At this stage only the 3rd family acquire a mass. 
The neutrino mass matrix and the first and second families masses arise
when we switch on the  explicitly breaking terms of $SU(3)$
$$
\delta \mathcal{L}_{A_4}={\bf 16}\,{\bf 16}\,\overline{\bf 126}_{s,t}+
{\bf 16}\,{\bf 16}\,{\bf 10}\,{\bf 45}'_{T_{3R}}\,{\bf 45}'_Y
$$
where  the scalar fields $\overline{\bf 126}_{s,t}$ are a singlet
${\bf 1'}$ and a triplet of $A_4$ respectively,  the
${\bf 45}'_{T_{3R}}$, ${\bf 45}'_Y$
are other scalars that transform as {\bf 45} of $SO(10)$, 
singlet and  triplet of $A_4$ respectively. 
The {\bf 10} is a singlet of $A_4$. It is not difficult to show that 
when type-II seesaw is dominant, the first term in
$\delta \mathcal{L}_{A_4}$ generates the light neutrino mass matrix 
of the form of  eq.~(\ref{15}). The second term in
$\delta \mathcal{L}_{A_4}$ gives a contribution like in 
eq.~(\ref{eq:L1}) and, after the 
breaking of $A_4$ into $Z_3$, it generates the first and second family masses, see eqs.~(\ref{break1})-(\ref{13}). 

\section{Conclusions}

In this paper we have proposed an embedding of the discrete $A_4$ flavor
symmetry in the larger continuous group $SO(3)_L\times SO(3)_R$ that 
explains in a natural way the huge hierarchy between the 3rd family  charged fermion masses and the others two. 
This is a consequence of the fact that $SO(3)_L\times SO(3)_R$ breaks 
spontaneously into $S_{3L}\times S_{3R}$ and gives a democratic mass
matrix that has only one massive eigenstate. If such eigenstate is assumed to be the 3rd family state, we still have an undeterminated
12 angle in the charged lepton sector that is fixed by breaking the 
democratic mass matrix.  
The crucial feature of our model is that once we break explicitly  $SO(3)_L\times SO(3)_R$ into $A_4$ we automatically generate  
first and second family charged fermion masses $m_{1,2}\ll m_3$. In order to fit the hierarchy between the masses of the first and second families, we require a tuning.
Assuming that the light neutrino Yukawa interactions come 
from the couplings with an $A_4$ singlet $\xi\sim {\bf 1'}$ and an $A_4$ 
triplet $\phi'$ that are scalar electroweak singlets and that
$\phi'$ acquires vev in the direction $(0,0,1)$,
we have showed that the lepton mixing matrix is the tribimaximal one.
The CKM is given by the identity matrix. Afterward we suggest how to generate the Cabibbo angle in the quark sector  through the introduction 
of higher order corrections. 
In particular in our model higher order operators give corrections of the same magnitude in each entries of all charged fermion mass matrices.
Assuming that the ratio between the  correction and $m_c$ is of the order
of the Cabibbo angle $\lambda$, we obtain that a rotation
of order $\lambda$ in the 12 plane appears in the up mass matrix.
However the down and charged lepton mass matrices are almost unaffected 
by corrections. This mismatching gives up to the Cabibbo angle.

Finally we have briefly discussed a $SO(10)$ realization of our model where the flavor group $SO(3)_L\times SO(3)_R$ is enlarged 
to $SU(3)$ and the democratic structure should arise  from the vev of a scalar that transform as a ${\bf \bar{6}}$ of the flavor group $SU(3)$.

\acknowledgments

Work supported by MEC grants FPA2005-01269 and FPA2005-25348-E, 
by Generalitat Valenciana ACOMP06/154. 

\bibliographystyle{h-elsevier.bst}
\bibliography{Left-Right-A4}

\end{document}